\documentclass[twocolumn,showpacs]{revtex4}
\usepackage{graphics,epsfig,graphicx}
\usepackage{color}
\usepackage{makeidx}
\usepackage{latexsym}
\usepackage{calrsfs}

\begin{document}

\title{Intensity-field correlation of single-atom resonance fluorescence }

\date{\today}

\author{S. Gerber$^1$, D. Rotter$^1$, L. Slodi\v{c}ka$^1$,
J. Eschner$^{1,4}$, H. J. Carmichael$^3$, and R. Blatt$^{1,2}$}

\affiliation{$^1$ Institute of Experimental Physics, University of
Innsbruck,
Technikerstr.\ 25, A-6020 Innsbruck, Austria \\
$^2$Institute of Quantum Optics and Quantum Information, Austrian Academy of Sciences, Innsbruck, Austria\\
$^3$Department of Physics, University of Auckland, Private Bag 92019, Auckland, New Zealand\\
$^4$ ICFO - Institut de Ciencies Fotoniques, Mediterranean Technology Park,
08860 Castelldefels (Barcelona), Spain
}


\begin{abstract}
We report measurements of an intensity-field correlation function
of the resonance fluorescence of a single trapped $^{138}$Ba$^+$
ion. Detection of a photon prepares the atom in its ground state
and we observe its subsequent evolution under interaction with a
laser field of well defined phase. We record the regression of the
resonance fluorescence source field. This provides a direct
measurement of the field of the radiating dipole of a single atom
and exhibits its strong non-classical behavior. In the setup an
interference measurement is conditioned on the detection of a
fluorescence photon.

\end{abstract}
\maketitle

Resonance fluorescence of atoms, in particular individual atoms,
has been the subject of quantum optical measurements for many
years \cite{Mandel, Loudon}. For example, resonance fluorescence
is routinely used as a tool to simply detect atoms, for
spectroscopy purposes, for creating photons, including for single
and twin-photon sources, and its quadrature components have been
used to create non-classical states of light \cite{Loudon}. In
short, the observation of resonance fluorescence is a technology
ubiquitous in experimental physics. While its features are well
investigated and understood, for spectroscopy and in quantum
optics, a direct observation of the time evolution of the source
field at the single-atom, single-photon level has not been made.

The measurement of correlation functions, sensitive to the
source-field, has been proposed in the seminal paper by W. Vogel
\cite{Vogel}. In a first and pioneering experiment, G. T. Foster
et al. have been able to report a wave-particle correlation
function of the field that emanates out of a cavity and
corresponds on average to only a fraction of a photon excitation
\cite{OroCar,CarOro,OroPRA}. Correlation functions of the fields
comprised of many photons have been observed with lasers and other
light sources \cite{Mandel,Loudon,Twiss2} and were approached
theoretically \cite{Laserphys}. However, to the best of our
knowledge, the time evolution of the field that corresponds to a
single resonance fluorescence photon has not been recorded so far.
Moreover, the only previous observation of the intensity-field
correlation \cite{OroCar,OroPRA} made use of a strong local
oscillator with photocurrent detection. The reported measurement
employs a weak local oscillator and photon counting and thus
operates in a different regime.

Motivation for this investigation is the development of a tool (i)
to investigate and monitor the emission of a single-photon field
and at a later stage (ii) to detect the influence of boundary
conditions, such as walls, mirrors, other atoms and quite
generally of a different (engineered) bath on the dynamics of the
emission process. For all of these tasks it will be necessary to
monitor the resonance fluorescence field and possibly even
feed-back \cite{Eschner,Bushev1,Bushev2,Uwe} on the radiating
dipole.

In this letter, we report on the measurement of a third-order
correlation function of a single radiating atom, using standard
photon counting techniques. Using a homodyne detection scheme, we
record the resonance fluorescence field conditioned on the
detection of an initial resonance fluorescence photon that
prepares the atom in its ground state. Since the correlation
function of two fields is termed $g^{(1)}$ and that of two
intensities is termed $g^{(2)}$ \cite{Twiss}, we accordingly coin
the name $g^{(1.5)}$ for this third-order intensity-field
correlation.

In this work the correlation measurement is triggered by detecting
a fluorescence photon from a single trapped $^{138}$Ba$^+$ ion,
which projects the ion into its ground state. Stop events are
obtained from a homodyne detector, where the fluorescence
interferes with a local oscillator (LO) of well-controlled phase
relative to the exciting laser. The experimental setup is
interferometrically stabilized and the phase of the LO can be
adjusted to anywhere within [0, 2$\pi$]. The measurement is
repeated many times for different phases of the LO- field, such
that the integrated signal records the average conditional time
evolution of the (fluctuating) amplitude of the electromagnetic
wave that constitutes the emission of a single resonance
fluorescence photon.

\begin{figure}[h!] \begin{center}
\includegraphics[width=8.75cm]{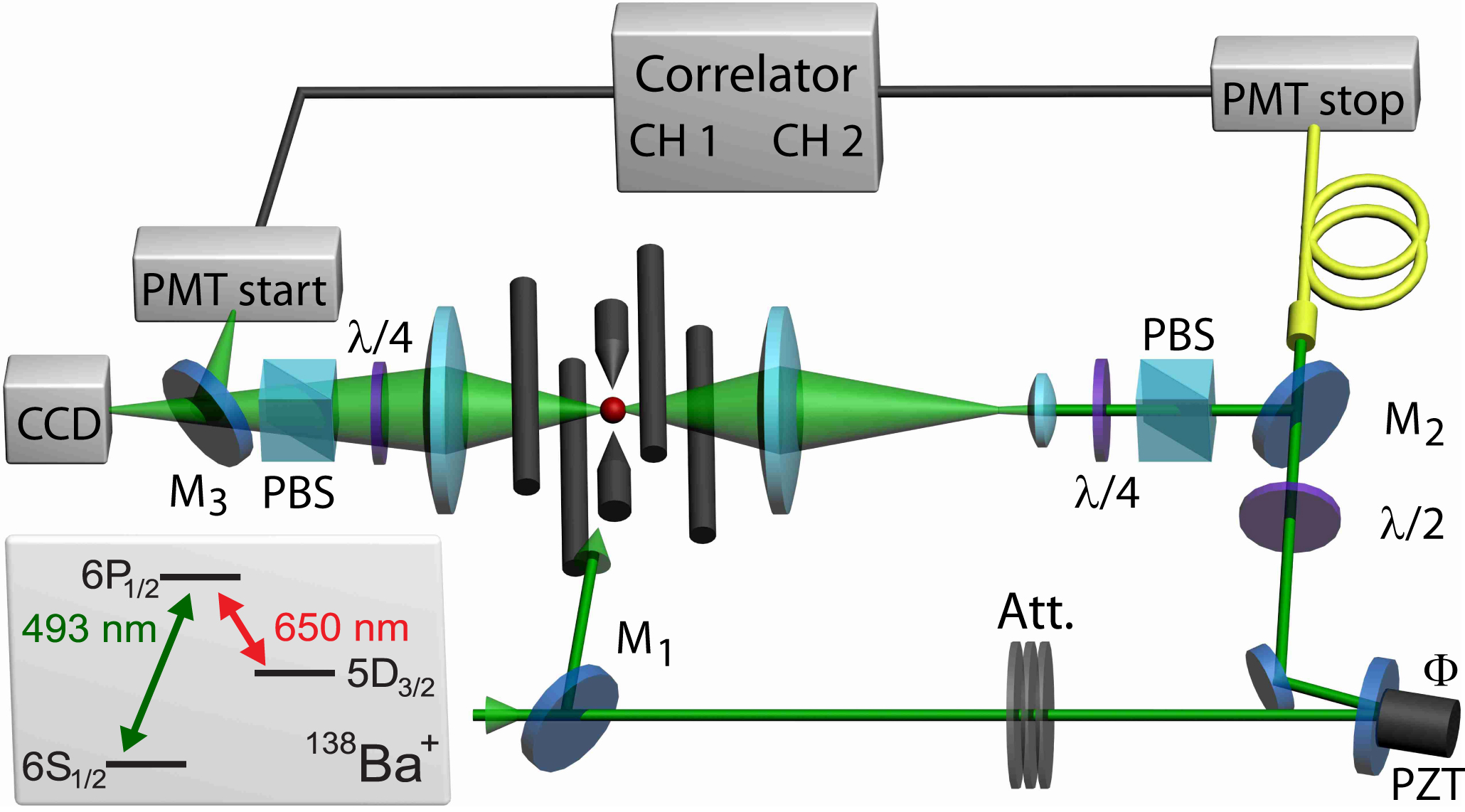} \caption{ (Color online)
Sketch of the experimental setup: A single
$^{138}$Ba$^{+}$ ion in a linear Paul trap is continuously laser excited.
Two detection channels, left and right, allow for visual observation of the ion (CCD),
or for recording correlations in the emitted light.
The LO is coupled out by $\rm M_1$ in front of the trap, attenuated (Att.),
and its polarization is adjusted with a $\lambda/2$ plate to match the polarization of
the fluorescence beam.
The inset shows the relevant
electronic levels of $^{138}$Ba$^{+}$. }
\label{fig1}
\end{center}
\end{figure}

The schematic experimental setup and the level scheme of the
$^{138}$Ba$^+$ ion are shown in Fig. \ref{fig1}. A single Ba$^+$
ion is loaded in a linear Paul trap using photo-ionization with
laser light near 413~nm \cite{Rotter_diss}. The ion is confined in
the harmonic pseudo-potential of the trap with radial (axial)
oscillation frequency $\sim$ 1.7~MHz ($\sim$\,1~MHz). Micromotion
is minimized using 3 pairs of dc electrodes. The ion is
continuously laser-cooled by two narrow-band (laser linewidth of a
few tens of kHz) linearly polarized tunable lasers at 493~nm
(green) and 650~nm (red) exciting the S$_{1/2}$--P$_{1/2}$ and
P$_{1/2}$--D$_{3/2}$ transitions, respectively. The green laser
intensity is adjusted to give mostly elastically scattered photons
\cite{Mandel}. After Doppler cooling, the ion is left in a thermal
motional state with a mean number of vibrational excitation
$<\widehat{n}>\,\approx15$. A weak magnetic field defines a
quantization axis perpendicular to the laser polarization
$\overrightarrow{E}$ and \overrightarrow{k} vector. Including the
Zeeman substates, the internal structure of the atom is described
as an 8-level system with the lasers exciting $\Delta m \pm 1$
transitions \cite{Toschek}.

Resonance fluorescence is detected in channels aligned along the
quantization axis in both directions. About 4$\%$ of the green
fluorescence is collected with two custom-made lenses (HALO
(LINOS), NA=0.4) placed about 1 cm from the trap center to the
left and right side of the trap.
The left beam can either be sent to a PMT (PMT-start) or to a CCD
camera. On the right hand side the fluorescence beam is collimated
with a telescope and then mixed with the LO field on a mirror $\rm
M_2$ with 99$\%$ reflectivity. After coupling to a single mode
optical fiber for mode matching, the interfering fields are
detected at another PMT (PMT-stop) leaving a count rate of about
10 kcps for the fluorescence after the fiber. In both detection
channels a quarter wave-plate and a Glan-Thompson polarizer select
$\Delta m =+1$ photons and filter out the $\Delta m =-1$
transition. The phase $\Phi$ of the interferometer is controlled
with a Piezo mounted mirror in the LO path by monitoring the count
rate of the homodyne signal. Thus the error in the phase of the LO
is given by the shot noise of this signal and is estimated to
about $\pm$10 degrees ($\phi=\pi/2$) and $\pm$24 degrees ($\phi=0$
or $\pi$) for a typical integration time of 0.1 s. Phase locking by
keeping the homodyne count rate constant is continuous with a time
constant of several seconds and does not affect the contrast of
our data within the limits set by the shot noise. Correlations
between the PMT start and the PMT stop-counts are obtained by
recording single photon arrival times with a Time Acquisition Card
(Correlator) with up to 100 ps resolution.

\begin{figure}[t] \begin{center}
\includegraphics[width=8.75cm]{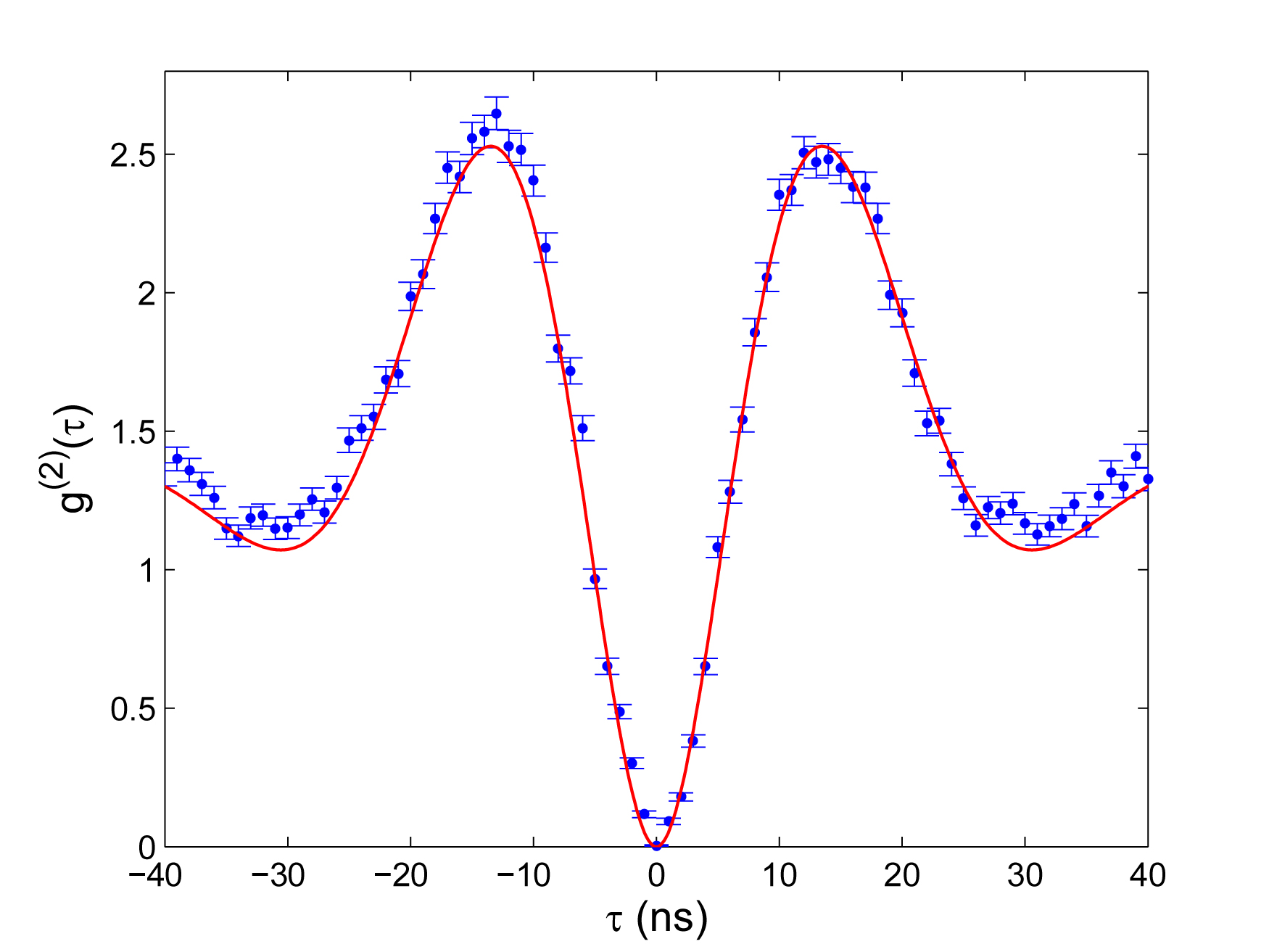} \caption{ (Color online)
Obtained correlation function with the LO path blocked.
The solid line shows the theoretical prediction
using experimental parameters and 8-level optical Bloch equations.
} \label{fig6}
\end{center}
\end{figure}

For a theoretical analysis we consider a frame rotating at the
(green) laser frequency, $\omega_L$. Thus, the green source part
of the radiated field by the ion reads
\begin{equation}
\widehat{E}(t)=\xi
e^{-i\omega_Lt}\hat\sigma^-(t),\label{eq1}
\end{equation}
where $t$ is in the long-time limit after the exciting laser is turned on,
$\xi$ represents a constant amplitude, and $\hat\sigma^-$ is the Pauli
lowering operator from $|P_{1/2},m=-1/2>$ to $|S_{1/2},m=+1/2>$,
associated with a creation of a single $\Delta m = +1$ photon. With
the LO path blocked we measure the conventional normalized second
order (intensity) correlation,
$g^{(2)}(\tau)\propto\langle\hat E^\dagger(0)\hat E^\dagger(\tau)\hat
E(\tau)\hat E(0)\rangle\equiv\lim_{t\to\infty}\langle\hat E^\dagger(t)\hat
E^\dagger(t+\tau)\hat E(t+\tau)\hat E(t)\rangle$.
In terms of atomic operators it reads
\begin{equation}
g^{(2)}(\tau)=\frac{\langle\hat\sigma^+(0)\hat\sigma^+
(\tau)\hat\sigma^-(\tau)\hat\sigma^-(0)\rangle}
{\langle\hat\sigma^+(0)\hat\sigma^-(0)\rangle^2}.\label{eq2}
\end{equation}
Figure \ref{fig6} depicts a measurement of this quantity. It
exhibits the characteristic anti-bunching at short time, with a
null rate of coincidences, $g^{(2)}(0)= 0.042(2)$ (without background subtraction).
Aside from this minor offset, it is well reproduced by our 8-level
Bloch simulations. Fitting parameters are the laser powers and detunings.
Thus, the $g^{(2)}(\tau)$ is used for calibrating the
laser settings for the $g^{(1.5)}(\tau)$ measurement.
With the LO arm unblocked, we measure the homodyne signal
conditioned on a photon emission from the ion, where the phase
$\Phi$ of the LO can be adjusted. We now write the detected fields
in units of the square root of photon flux.
$\gamma_1\langle\hat\sigma^+\hat\sigma^-\rangle$ represents the
mean photon flux into the PMT-start, where $\gamma_1$ is the
product of the radiative decay rate and the overall collection and
detection efficiency of the PMT-start. We similarly denote the
fluorescence field at the PMT-stop by
$\sqrt{\gamma_2}\hat\sigma^-(0)$, where $\gamma_2$ is the product
of radiative decay rate and the respective collection and
detection efficiency of a photon at the PMT-stop. Then
representing the local oscillator field by the complex amplitude
$\mathcal{E} e^{i\Phi}$ the field after the interferometer reads
\begin{equation}
X_\Phi(t)=[\mathcal{E} e^{i\Phi}+ \sqrt{\gamma_2}\hat\sigma^-(t)],
\end{equation}
and for positive $\tau$ we measure a total unnormalized second-order correlation
\begin{equation}
G^{\rm total}_\Phi(t,t+\tau)=\langle\sqrt{\gamma_1}\hat\sigma^+(t)X_{\Phi}^\dag(t+\tau)
X_{\Phi}(t+\tau)\sqrt{\gamma_1}\hat\sigma^-(t)\rangle, \label{eq4}
\end{equation}
which expands out to
\begin{equation}
G^{\rm total}_\Phi(\tau)=F\big\{(1-V)[(1-r)+rg^{(2)}(\tau)]+V g^{(1.5)}_\Phi(\tau)\big\}
.\label{eq5}
\end{equation}
Here, $g^{(2)}(\tau)$ is the intensity correlation
function given by Eq.\,(\ref{eq2}),
and for the third-order correlation function at a given LO phase we write
\begin{equation}
g^{(1.5)}_\Phi(\tau)=\frac{\langle
\hat\sigma^+(0)[e^{i\Phi}\hat\sigma^+(\tau)+e^{-i\Phi}\hat\sigma^-(\tau)]\hat\sigma^-(0)\rangle}
{\langle\hat\sigma^+
+\hat\sigma^-\rangle\langle\hat\sigma^-\hat\sigma^+\rangle }.
\label{eq6}
\end{equation}
For later convenience we define the LO phase relative to the asymptotic phase of the resonance
fluorescence field,
such that $g^{(1.5)}_{\pi/2}(\tau \rightarrow \infty)=0$. The pre-factor $F$ in
Eq.\,(\ref{eq5}) is
\begin{equation}
F=\gamma_1\langle\hat\sigma^+\hat\sigma^-\rangle(\mathcal{E}^2 +\mathcal{E}\sqrt{\gamma_2}
\langle\hat\sigma^+ +\hat\sigma^-\rangle+  \gamma_2 \langle\hat\sigma^+ \hat\sigma^-\rangle),
\end{equation}
while
\begin{equation}
V=\frac{\mathcal{E}\sqrt{\gamma_2}\langle\hat\sigma^+ +\hat\sigma^-\rangle}
{\mathcal{E}^2+\mathcal{E}\sqrt{\gamma_2}\langle\hat\sigma^+ +\hat\sigma^-\rangle+\gamma_2
\langle\hat\sigma^+ \hat\sigma^-\rangle}
\end{equation}
is the visibility of the interference part in $G^{\rm total}_\Phi(\tau)$ and
\begin{equation}
r=\frac{\gamma_2\langle\hat\sigma^+\hat\sigma^-\rangle}{\mathcal{E}^2+\gamma_2\langle\hat\sigma^+\hat\sigma^-\rangle}\\
\end{equation}
is the ratio of the florescence intensity to the total intensity at the PMT-stop.

According to Eq.\,(\ref{eq5}), the expected correlation function
consists of three parts. A $\Phi$-dependent part with visibility
$V$ reveals the $g^{(1.5)}(\tau)$ correlation due to the
interference of the fluorescence with the LO. The remaining
non-interfering part with weight $1-V$ consists of a nor- mal
second-order correlation function $g^{(2)}(\tau)$ (both start and
stop counts from fluorescence photons) weighted by $r$ and a
constant offset (stop counts from LO photons) weighted by $1-r$.
Normalizing $G^{\rm total}_\Phi(\tau)$ by $F(1-V)$ we obtain
\begin{eqnarray}
g^{\rm total}_\Phi(\tau)=(1-r)+rg^{(2)}(\tau)
+\frac{V}{1-V}g^{(1.5)}_\Phi(\tau).\label{eq10}
\end{eqnarray}
This normalization is chosen such that at a phase $\Phi=\pi/2$ of the LO,
when $g^{(1.5)}(\tau)$ vanishes
asymptotically, $g^{\rm total}(\tau)$ yields an asymptotic value of 1.

\begin{figure}[t] \begin{center}
\includegraphics[width=8.75cm]{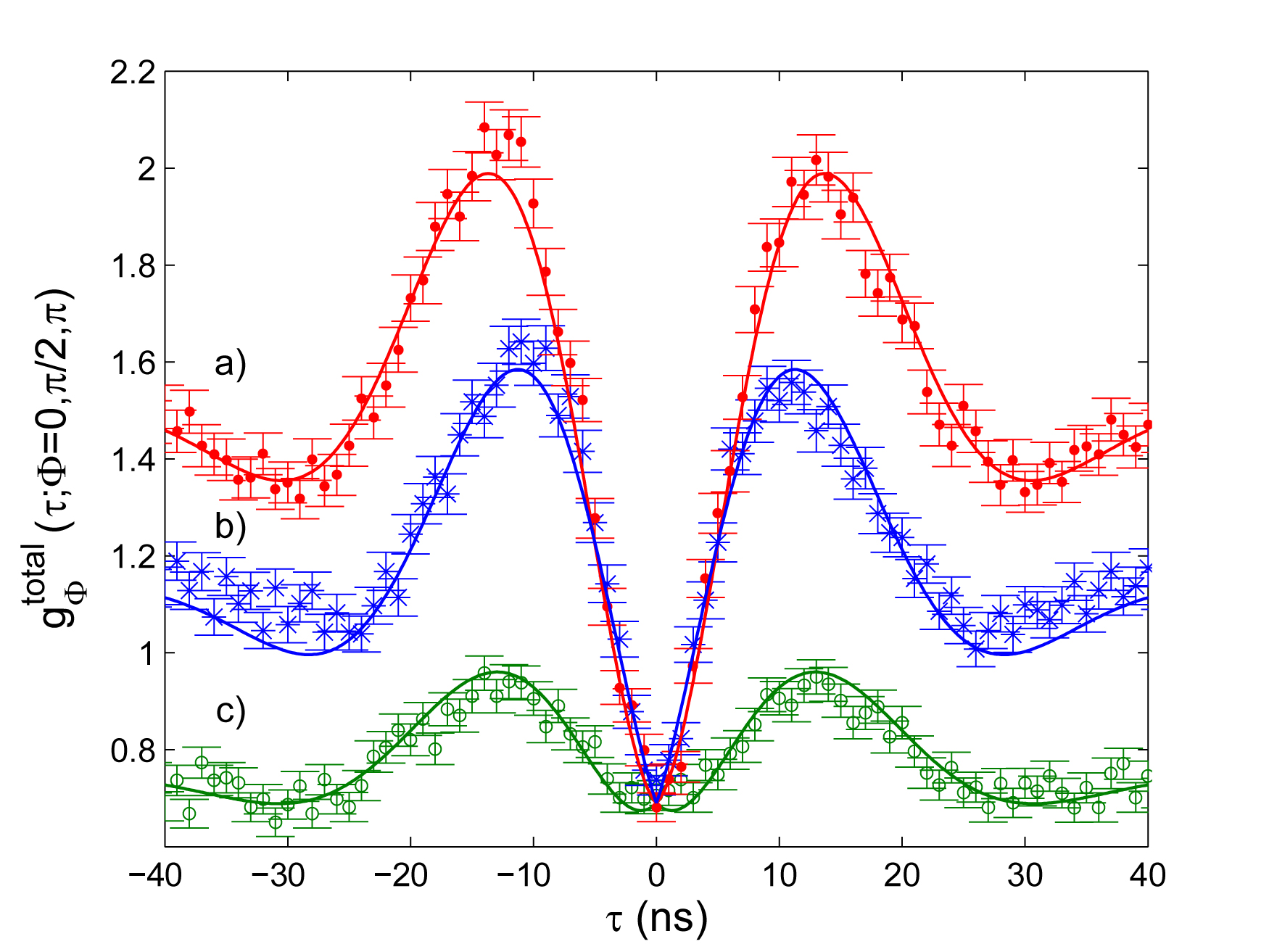} \caption{ (Color online)
Measured normalized correlation function $g^{\rm
total}_\Phi(\tau)$ obtained a) at $\Phi=0$; i.e. the fluorescence
being in-phase with the LO , c) at  $\Phi=\pi$; i.e. the
fluorescence being out-of-phase with the LO and b) at
$\Phi=\pi/2$. The solid curves show the theoretical predictions
using Eq.\,(\ref{eq10}). } \label{fig2}
\end{center}
\end{figure}

Figure \ref{fig2} shows the measured correlations between
PMT-start and PMT-stop with the LO phase adjusted to
$\Phi=0,\pi/2$ and $\pi$. Data are acquired after 30~minutes of
accumulation for each curve and presented with a 1~ns resolution.
The corresponding variance is obtained from shot noise, i.e.
assuming Poisson statistics at all times $\tau$. The solid curves
show the theoretical prediction using Eq.\,(\ref{eq10}) with a
visibility $V\sim18\%$ and an intensity ratio $r=0.31$. The
measured correlations are well reproduced by superposition of the
three contributions described by Eq.\,(\ref{eq10}).
All curves contain a constant contribution and a scaled
$g^{(2)}(\tau)$ correlation. In addition, curve b), where the LO
phase is set to $\Phi=\pi/2$, contains the imaginary part of the
atomic polarization whose asymptotic contribution is zero. In
contrast, curves a) and c), where the LO phase is set to 0 and
$\pi$, respectively, reveal the real part of the polarization
which adds constructively or destructively to the other two
contributions. All curves show the same coincidence rate at
$\tau=0$. Since both the $g^{(2)}(\tau)$ contribution and the
$g^{(1.5)}_\Phi(\tau)$ contribution are identically zero at
$\tau=0$, the measured coincidence rate at this point is solely
determined by the offset of non- interfering LO photons at
PMT-stop (and background counts). Calibration of the LO phase is
obtained by looking for the maximum and minimum asymptotic values
of $G^{\rm total}_\Phi(\tau)$ and assigning to them the LO phases
$\Phi=0$ and $\pi$, respectively.

\begin{figure}[t] \begin{center}
\includegraphics[width=8.75cm]{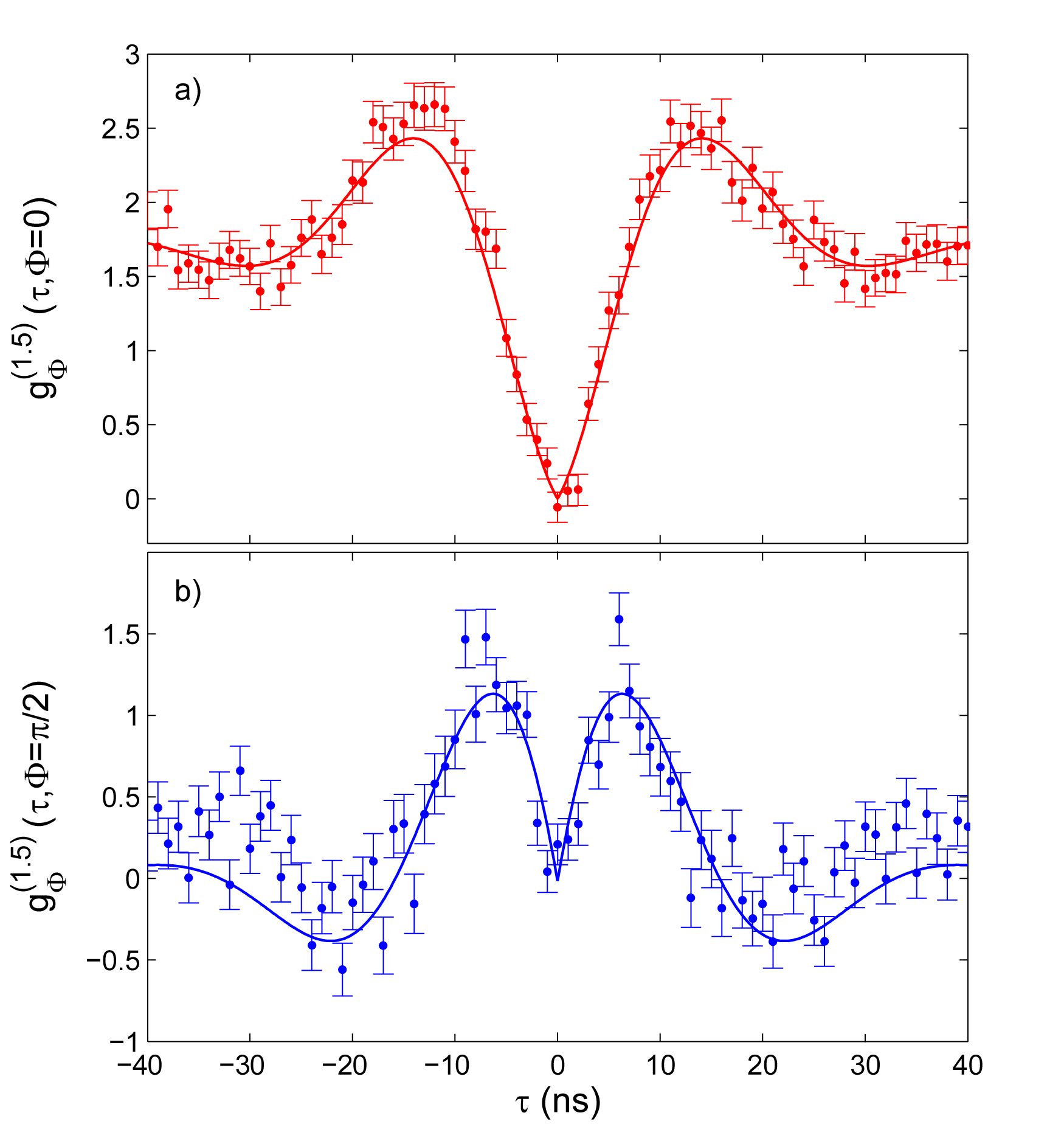}
\caption{(Color online)
The intensity-field correlation function $g^{(1.5)}_\Phi(\tau)$
with the LO phase adjusted to a) $\Phi=0$ and b) $\Phi=\pi/2$,
obtained from the data in Fig. \ref{fig2} using  Eq.\,(\ref{eq8})
and Eq.\,(\ref{eq11}),  respectively. The solid lines represent
the theoretical model using Eq.\,(\ref{eq6}). } \label{fig4}
\end{center}
\end{figure}

Determining the full complex $g^{(1.5)}(\tau)$ intensity field
correlation function requires its measurement for two orthogonal
phases. We deduce it from the data in Fig. \ref{fig2} and using
Eq.\,(\ref{eq10}) in the following way:
\begin{equation}
g^{(1.5)}_0(\tau)=\frac{1-V}{2V}( g^{\rm total}_0(\tau)-g^{\rm total}_\pi(\tau))
\label{eq8}
\end{equation}
and
\begin{equation}
g^{(1.5)}_{\pi/2}(\tau)=\frac{1-V}{2V}\Big( 2g^{\rm total}_{\pi/2}(\tau)-
(g^{\rm total}_0(\tau)+g^{\rm total}_\pi(\tau))\Big).
\label{eq11}
\end{equation}
The result is shown in Fig. \ref{fig4} together with the
theoretical prediction of Eq.\,(\ref{eq10}) (solid line). The data
reveal the time evolution of the fluorescence field as it evolves
from its initialization by an emitted photon into steady-state via
damped Rabi oscillations. Comparing Fig. \ref{fig4} a) with Fig.
\ref{fig6} we see that the $g^{(1.5)}(\tau)$ grows linearly with
$\tau$ around the point $\tau=0$ while the $g^{(2)}(\tau)$ grows
quadratically with $\tau$ showing clearly that the field rather
than the intensity is measured.
The limitation for the visibility in the homodyne part of the
setup is determined by the temporal overlap of two photon wave
packets impinging at the mixing mirror. It is limited by the
coherence time and the flux of the fluorescence photons with
respect to the LO photons. Assuming elastically scattered photons,
the coherence time is given by the 493~nm laser bandwidth with
$T=1/\Delta \nu\approx\,$50$\,\mu s$ \cite{Raab}. The fluorescence
count-rate of $10$\,kcps is predetermined by the collected
fraction of solid angle and the fiber-coupling efficiency. While
the temporal overlap would benefit from a higher LO intensity
(smaller $r$), the visibility of the interferometer would decrease
and locking the interferometer would get more involved due to a
larger shot-noise in the LO arm. Thus, optimizing the experiment
resulted in a reduced visibility of $V\sim18\,\%$ and an intensity
ratio of $r=0.31$.

In summary, we have successfully measured an intensity-field
correlation function of the resonance fluorescence from a single
$^{138}$Ba$^+$ ion. In our setup a photon detection from the ion
starts the correlation measurement in a well defined state that
evolves to steady-state via Rabi oscillations. The correlation
function $g^{(1.5)}$ was obtained from the measured data points
recorded with the LO being in and out-of phase with the
fluorescence.

This measurement clearly shows the dynamical behavior of the
atomic dipole. In principle, these measurements now allow for a
detailed investigation of the fluctuating dipole and its
non-classical statistics that leads to the fact that resonance
fluorescence produces inherently squeezed light \cite{Mand}. It is
possible in principle to observe this effect using the third-order
correlation function \cite{Vogel,OroPRA}. However, for this, the
single atom must be only weakly excited which was not the case in
the present experiment. Observation of the squeezing of the
single-ion resonance fluorescence will be subject to further
investigations. Finally, the current procedure will allow us to
investigate the radiating dipole field under the influence of
direct backaction \cite{Eschner} and in the presence of boundary
conditions \cite{Uwe,Dubin} and active feedback \cite{Bushev2}.

We thank L. Orozco for valuable discussions. This work has been
partially supported by the Austrian Science Fund FWF (Project No.
SFB F-015) and by the Institut f$\ddot{\rm u}$r Quanteninformation
GmbH. J.E. acknowledges support by the European Commission
("EMALI", MRTN-CT-2006-035369).

\end{document}